\documentstyle[12pt]{article}
\renewcommand
\baselinestretch{1.4}

\def\beq{\begin{equation}}
\def\brr{\begin{array}}
\def\err{\end{array}}
\def\eeq{\end{equation}}
\def\bea{\begin{eqnarray}}
\def\eea{\end{eqnarray}}
\def\bs{\bigskip}
\def\tr{\mbox{Tr}\, }
\def\ni{\noindent}
\def\wt{\widetilde}

\def\nn{\nonumber}
\def\ms{\medskip}

\def\re{\mbox{Re}\, }

\begin{document}

\hfill HUPD-92-12

\hfill UB-ECM-PF 92/25

\hfill UTF 273-92

\hfill September 1992

\vspace*{3mm}

\begin{center}

{\LARGE \bf
A novel representation for the free energy in string theory at
non-zero temperature}

\vspace{4mm}

\renewcommand
\baselinestretch{0.8}

{\sc A.A. Bytsenko} \\ {\it Physico-Mechanical Faculty, State
Technical
University, St Petersburg 195251, Russia} \\
{\sc E. Elizalde}\footnote{E-mail address: eli @ ebubecm1.bitnet}
\\
{\it Department E.C.M., Faculty of Physics, University of
Barcelona, \\
Diagonal 647, 08028 Barcelona, Spain} \\
{\sc S.D. Odintsov}\footnote{On sabbatical leave from
Tomsk Pedagogical Institute, 634041 Tomsk, Russia.} \\
{\it Department of Physics, Faculty of Science, Hiroshima
University, \\
Higashi-Hiroshima 724, Japan} \\
 {\sc S. Zerbini} \\
{\it Department of Physics, University of Trento, 38050 Povo,
Italy} \\
and      \\
{\it I.N.F.N., Gruppo Collegato di Trento}
\ms

\renewcommand
\baselinestretch{1.4}

\vspace{5mm}

{\bf Abstract}

\end{center}

A novel representation ---in terms of a Laurent series--- for the free
energy of string theory at non-zero temperature is constructed. The
examples of open bosonic, open supersymmetric and closed bosonic
strings are studied in detail. In all these cases the Laurent
series representation for the free energy is obtained explicitly.
 It is shown that the Hagedorn temperature arises in
this
formalism as the convergence condition (specifically, the radius of
convergence) of the corresponding Laurent series. Some prospects
for further applications are also discussed. In particular, an
attempt to describe string theory above the Hagedorn temperature
---via Borel analytical continuation of the Laurent series
representation--- is provided.



\newpage

\section{Introduction}

There is a number of motivations for the study of string theory at
non-zero temperature (for a short list of references see [1-6]; a
general review with a more complete list can be found in [5]). One
of the main reasons for its study is connected with the fact that
string thermodynamics can be relevant for the description of the
very early universe (see [7] and references therein). This will be,
of course, the case if (finally) the theory of everything can be
constructed on a stringy foundation. Another (recent) motivation is
related with the attempts [8,9] to use string theory at non-zero
temperature  for the description of the high temperature limit of
the confining phase of large-N SU(N) Yang-Mills theory.

It is well known since the early days of dual string models, that an
essential ingredient of string theory at non-zero temperature is
the so-called Hagedorn temperature, above which the free energy
diverges.
The appearance of the Hagedorn temperature is a consequence of the
fact that the asymptotic form of the state level density has an
exponential dependence of the mass.
According to the popular viewpoint, the Hagedorn temperature
is the critical temperature for a phase transition to some new
phase (probably topological string theory [10]), which describes
string
theory above the Hagedorn temperature. Unfortunately, at present
there is almost nothing to be said about this new phase of string
theory. In  such a situation, it is reasonable to
make an additional effort in order to try to understand in more
depth
string thermodynamics below the Hagedorn temperature, before
becoming
involved in the phase transition itself.

There are different representations (in particular, those connected
with different ensembles) for the string free energy. One of these
representations [6] ---which is very useful for formal
manipulations--- gives a modular-invariant expression for the free
energy. However, that and all the other well known representations
(see [5]) are {\it integral} representations, in which the Hagedorn
temperature appears as the  convergence condition in the
ultraviolet limit of a certain integral. In order to
discuss the high- or the low-temperature limits in such
representations one must expand the integral in terms of a
corresponding series. Thus, a specific series expansion
appears in string theory at non-zero temperature.

In the present article we develop a formalism ---steming from
field theory at non-zero temperature [11,12]--- for the description
of string theory at non-zero temperature. Within this formalism, we
construct a novel representation for the (super)string free energy,
namely one in terms of a Laurent series.  This representation
is very convenient, both at high and at low  temperatures. What is
most interesting is the fact that
the Hagedorn temperature appears in this formalism as the radius of
convergence of the Laurent series.
The paper is organized as follows. We begin, in section 2, with the
description of the Laurent series representation for the free
energy in field theory at non-zero temperature. In section 3 we
apply this formalism to open string theory at non-zero temperature
(both to open bosonic strings and to open superstrings). Section 4
is devoted to the construction of a Laurent series representation
for the closed bosonic string. In section 5 a discussion of the
results is presented. In Appendix A we carry out a simple
derivation of the Hagedorn temperature using the heat kernel
method. Appendix B is devoted to an attempt at the description of
the free energy above the Hagedorn temperature (i.e., in the new
phase) via the Borel analytical continuation of our Laurent series.
Appendix C contains the explicit high-temperature expansion of the
free energy for open (super)strings.
\bs

\section{Free energy in field theory at non-zero temperature}

 It is  well-known that the one-loop
free energy for the bosonic ($b$) or fermionic ($f$) degree of
freedom in $d$-dimensional space is given by
\beq
F_{b,f} = \pm \frac{1}{\beta} \int \frac{d^{d-1} k}{(2\pi )^{d-1}}
\log \left(1\mp  e^{-\beta u_k} \right)
\eeq
where $\beta$ is the inverse temperature, $u_k=\sqrt{k^2+m^2}$, and
$m$ is the mass for the corresponding degree of freedom.
Expanding the logarithm and performing the (elementary) integration
one easily gets (see for example [13])
\bea
F_b &=& - \sum_{n=1}^{\infty} (\beta n)^{-d/2} \pi^{-d/2} 2^{1-d/2}
m^{d/2} K_{d/2} (\beta n m), \nn \\
F_f &=& - \sum_{n=1}^{\infty} (-1)^n (\beta n)^{-d/2} \pi^{-d/2}
2^{1-d/2} m^{d/2} K_{d/2} (\beta n m),
\eea
where $ K_{d/2} (z)$ are the modified Bessel functions. Using the
integral representation for the Bessel function
\beq
 K_{d/2} (z) = \frac{1}{2} \left( \frac{z}{2} \right)^{d/2}
\int_0^{\infty} ds \, s^{-1-d/2} e^{-s-z^2/(4s)},
\eeq
one can obtain the well-known proper time representation for the
one-loop free energy:
\beq
F_{b,f} = -
\pi^{-d/2} 2^{-1-d/2}
\int_0^{\infty}   ds \,
s^{-1-d/2}
e^{-m^2s/2} \times \left\{ \brr{l} \left[ \theta_3  \left( 0 |
\frac{i\beta^2}{2\pi s} \right) -1 \right], \\
\left[ 1-\theta_4  \left( 0 | \frac{i\beta^2}{2\pi s} \right)
\right].  \err \right.
\eeq
Expression (4) is usually the starting point  for the calculation
of the string free energy in the canonical ensemble (then $m^2$ is
the mass operator and for closed strings the corresponding
constraint  should be taken into account).
Let us now consider a different representation (in terms of a
Laurent series) for the bosonic or fermionic field free energy in
a
$d$-dimensional space-time. In field theory such a representation
has been introduced in refs. [11,12] (see also [14], where
a representation
involving $\beta^{-1} \log Z$ had already been given, but in a
particular case only).
We start from the obvious identity
\beq
\int d^d k \, f(k) = \frac{2\pi^{d/2}}{\Gamma (d/2)} \int dk \,
k^{d-1} f(k).
\eeq
Using it and upon integration of (1) by parts, we obtain
\beq
F_{b,f} =- \frac{(4\pi)^{(1-d)/2}}{(d-1)\Gamma ((d-1)/2)} \int dk^2
\,  \frac{k^{d-1}}{u_k \left( e^{\beta u_k} \mp 1 \right)}.
\eeq
For the factor in the integrand, $\left( e^{\beta u_k} \mp 1
\right)^{-1}$, we shall use the Mellin transform in the following
form [15]
\beq
 \frac{1}{ e^{ax} \mp 1 } = \frac{1}{2\pi i} \int_{c-
i\infty}^{c+i\infty} ds \,  \zeta^{(\mp)} (s) \Gamma (s)
(ax)^{-s},     \ \  \ \ \ \ ax >0,
\eeq
where Re $s=c, \ c>1$, for bosons and $c>0$ for fermions, $
\zeta^{(-)} (s)=\zeta (s)$ is the Riemann-Hurwitz zeta function and
$ \zeta^{(+)} (s)=\left( 1-2^{1-s} \right) \zeta (s)$ $=
\sum_{n=1}^{\infty} (-1)^{n-1} n^{-s}$, Re $s>0$.

Substituting (7) into (6), we get
\beq
F_{b,f} =- \frac{(4\pi)^{(1-d)/2}}{(d-1)\Gamma ((d-1)/2)} \int dk^2
\, k^{d-1} \frac{1}{2\pi i} \int_{c-i\infty}^{c+i\infty} ds \,
\zeta^{(\mp)} (s) \Gamma (s) \beta^{-s} u_k^{-1-s}.
\eeq
Integrating over $k$ with the help of the Euler beta function
$B(x,y)=\Gamma (x)\Gamma (y)/\Gamma (x+y)$ (notice that, owing to
absolute convergence, the order of integration over $k$ and $s$ can
be interchanged), we obtain
\beq
 \int dk^2 \, k^{d-1}  u_k^{-1-s} = (m^2)^{(d-s)/2} B \left(
\frac{d+1}{2}, \frac{s-d}{2} \right) = (m^2)^{(d-s)/2} \frac{
\Gamma \left( \frac{d+1}{2} \right) \Gamma \left( \frac{s-d}{2}
\right)}{ \Gamma \left( \frac{s+1}{2} \right)}, \ \ \mbox{Re} \, s
>d.
\eeq
Finally, one gets [11]
\beq
F_{b,f} =- 2^{-d} \pi^{(1-d)/2} \frac{1}{2\pi i} \int_{c-
i\infty}^{c+i\infty} ds \,  \Gamma (s) \frac{ \Gamma \left(
\frac{s-d}{2} \right)}{ \Gamma \left( \frac{s+1}{2} \right)}
\zeta^{(\mp)} (s)  \beta^{-s}  (m^2)^{(d-s)/2},  \ \ \
c >d.
\eeq
This is the main expression which will be used for the calculation
of the one-loop string free energy. Notice that the new formula
(10)
is quite different from the usual one (4).

For string theory one can write the representation (10) in the form
\bea
F_{\mbox{bosonic string}} &=&- 2^{-d} \pi^{(1-d)/2} \frac{1}{2\pi
i} \int_{c-i\infty}^{c+i\infty} ds \,  \Gamma (s) \frac{
\Gamma \left( \frac{s-d}{2} \right)}{ \Gamma \left( \frac{s+1}{2}
\right)} \zeta (s)  \beta^{-s} \tr (M^2)^{(d-s)/2}, \nn \\  & &
d=26,
\ \ \ c>d,
 \\
F_{\mbox{superstring}} &=&- 2^{-d+1} \pi^{(1-d)/2} \frac{1}{2\pi i}
\int_{c-i\infty}^{c+i\infty} ds \,  \Gamma (s) \frac{ \Gamma
\left( \frac{s-d}{2} \right)}{ \Gamma \left( \frac{s+1}{2} \right)}
\zeta (s)  (1-2^{-s}) \beta^{-s} \nn \\
&& \times \mbox{Str} \, (M^2)^{(d-s)/2}, \ \ \ \ \
d=10,
\ \ \ c>d.
\nn  \eea
Here $M^2$ is the mass operator and for closed strings the
constraint should be introduced via the usual identity [2]. It is
interesting to observe that the fermionic contribution to the free
energy can be obtained from the bosonic one with the help of (10).
The closed result is [16]
\beq
F_f (\beta ) = - \left( 2F_b (2\beta ) -F_b (\beta ) \right).
\eeq
\bs

\section{Laurent series representation for the open (super)string
free energy}

For bosonic strings the mass operator contains both infrared (due
to the presence of the tachyon in the spectrum) and ultraviolet
divergences, while for superstrings it contains only ultraviolet
divergences. Hence, the consideration of superstrings will be much
simpler from a technical point of view.

In what follows we shall consider open bosonic strings and open
superstrings.  For the open superstring (without gauge
group) the spectrum is given by (see for example [17])
\beq
M^2=2\sum_{i=1}^{d-2}\sum_{n=1}^{\infty}  n \left( N_{ni}^b +
N_{ni}^f \right), \ \ \ \ d=10.
\eeq
In order to calculate Str $(M^2)^{(d-s)/2}$ it is convenient to use
the heat-kernel representation
\beq
\mbox{Str} \, (M^2)^{(d-s)/2}= \frac{1}{\Gamma \left( \frac{s-d}{2}
\right)} \int_0^{\infty} dt \, t^{ \frac{s-d}{2} -1} \, \mbox{Str}
\,
e^{-tM^2}.
\eeq
For $d=10$, it is known that
\beq
 \mbox{Str} \, e^{-tM^2} =\prod_{n=1}^{\infty} \left( \frac{1-e^{-
2tn}}{1+e^{-2tn}} \right)^{-8} = \left[ \theta_4 \left( 0 |
e^{-2t} \right)\right]^{-8},
\eeq
and, substituting,
\beq
\left. \mbox{Str} \, (M^2)^{ \frac{d-s}{2}}\right|_{d=10}  =
\frac{2^{5-s/2}}{\Gamma \left( \frac{s}{2}-5 \right)}
\int_0^{\infty} dt \, t^{ \frac{s}{2} -6} \left[ \theta_4 \left( 0
| e^{-t} \right)\right]^{-8}.
\eeq
According to (10), Re $s>10$, this is why, for $s\leq 10$,
expression (16) contains  ultraviolet divergences ($t\rightarrow
0$).

Let us now analytically continue the integral (16) to the complex
$s$-plane. Recall that
\bea
\theta_4 \left( 0 | e^{-t} \right) &=&
\sqrt{\frac{\pi}{t}} \,
\theta_2 \left( 0 | e^{-\pi^2/t} \right) =
\sqrt{\frac{\pi}{t}}
\sum_{n=-\infty}^{+\infty} \exp \left[ -\frac{\pi^2}{t} \left( n-
\frac{1}{2} \right)^2 \right] \nn \\
&=& 2
\sqrt{\frac{\pi}{t}}
 \, e^{-\pi^2/(4t)} \left( 1+ e^{-9\pi^2/t}+ e^{-25\pi^2/(4t)} +
\cdots \right), \ \ \ \ t\longrightarrow 0,
\eea
and therefore
\beq
\left. \left[ \theta_4 \left( 0 | e^{-t} \right)\right]^{-
8} \right|_{t\rightarrow 0} = \frac{t^4}{2^8\pi^4} \, e^{2\pi^2/t}
-\frac{t^4}{2^5\pi^4} + {\cal O} \left( e^{-2\pi^2/t} \right).
\eeq
Hence, one can identically regularize the integral (16) in the
following way
\bea
 \mbox{Str} \, (M^2)^{5- \frac{s}{2}}& =&  \frac{2^{5-s/2}}{\Gamma
\left( \frac{s}{2}-5 \right)} \left\{ \int_0^{\infty} dt \, t^{
\frac{s}{2} -6} \left[ \left[ \theta_4 \left( 0 | e^{-t}
\right)\right]^{-8} - \frac{t^4}{2^8\pi^4} \, \left( e^{2\pi^2/t}
-
8 \right) \right] \right. \nn \\
& +&\left.   \frac{1}{2^8\pi^4} \, \int_0^{\infty} dt \, t^{
\frac{s}{2} -2} \left( e^{2\pi^2/t} -8 \right)
\right\}.
\eea
Since the regularization of the integral $ \int_0^{\infty} dx \,
x^{\lambda}$ as an analytical function of $\lambda$ gives $
\int_0^{\infty} dx \, x^{\lambda} =0$, it turns out that the last
integral in (19) is equal to zero. Moreover, we have
\beq
 \int_0^{\infty} dt \, t^{ \frac{s}{2} -6} t^4 e^{2\pi^2/t}
=   \int_0^{\infty} dt \, t^{\left(- \frac{s}{2}
+1\right)-1}  e^{2\pi^2t} =(- 2\pi^2)^{ \frac{s}{2} -1} \Gamma
\left(1-\frac{s}{2} \right),
\eeq
and therefore
\beq
 \mbox{Str} \, (M^2)^{5- \frac{s}{2}} = \frac{2^{-4}\pi^{-
6}}{\Gamma \left(
\frac{s}{2}-5 \right)} \left[ \pi^s \Gamma \left(1-\frac{s}{2}
\right) \mbox{Re} \,  (-1)^{\frac{s}{2} -1}  +2 G(s,\mu)
\right],
\eeq
where
\beq
G(s, \mu)= \pi \left( \frac{\pi}{2} \right)^{s/2}  \int_0^{\mu} dt
\, t^{ \frac{s}{2} -6}
\left\{ \left[ \frac{1}{2} \theta_4 \left( 0 | e^{-\pi t}
\right)\right]^{- 8} -  t^4\left( e^{2\pi/t}-8 \right) \right\}.
\eeq
In (22) the infrared cutoff parameter $\mu$  has been introduced.
Such a regularization is necessary  for $s\geq 4$. On the next
stage of our calculations this regularization will be removed ($\mu
\rightarrow \infty$).

For the one-loop free energy, we obtain
\bea
F_{\mbox{superstring}} &=& - 2^{-13} \pi^{-21/2} \frac{1}{2\pi i}
\int_{c-i\infty}^{c+i\infty} ds \, \frac{ \Gamma
( s)}{ \Gamma \left( \frac{s+1}{2} \right)} \zeta (s) (1-2^{-s})
\beta^{-s}\nn \\
&\times& \left[ \pi^s \Gamma \left(1-\frac{s}{2} \right) \mbox{Re}
\,  (-1)^{\frac{s}{2} -1}  +2 G(s,\mu)
\right] \\
& \equiv &  - 2^{-13} \pi^{-21/2} \frac{1}{2\pi i}
\int_{c-i\infty}^{c+i\infty} ds \, \left[ \varphi (s) + \psi (s)
\right], \ \ \ \ c>10. \nn
\eea
Here
\bea
\varphi (s) &=& (1-2^{-s} )  \mbox{Re} \,  (-1)^{\frac{s}{2} -1}
\frac{ \Gamma ( s) \Gamma \left(1- \frac{s}{2} \right)}{ \Gamma
\left( \frac{s+1}{2} \right)} \zeta (s) \left( \frac{\beta}{\pi}
\right)^{-s}, \\
\psi (s) &=& 2 (1-2^{-s} )  G(s,\mu) \,  \frac{ \Gamma ( s)}{
\Gamma \left( \frac{s+1}{2} \right)} \zeta (s)  \beta^{-s}.
\eea

The meromorphic function $\varphi (s)$ has first order poles at
$s=1$ and $s=2k$, $k=1,2, \ldots$. The meromorphic function $\psi
(s)$ has first order poles at $s=1$. The corresponding residues are
\beq
\mbox{Res} \, (\varphi (s),s=2k)=  (1-2^{-2k}) \frac{ \Gamma
(2k)\zeta (2k)}{ \Gamma \left(k+ \frac{1}{2} \right) \Gamma (k)}
\left( \frac{\beta}{\pi} \right)^{-2k}
\eeq
and
\beq
\mbox{Res} \, (\psi (s),s=1)=   G(1, \infty) \beta^{-1}.
\eeq
The pole of $\varphi$ for $s=1$  is also of first order but its
residue
is imaginary. There
are no
further poles contributing to the integral. One can see that the
regularization cutoff in the infrared domain is removed
automatically.

Finally, we obtain
\bea
F_{\mbox{superstring}} & =& -2^{-13} \pi^{-21/2} \left[ \sum_k
\mbox{Res} \, (\varphi (s),s=2k) +\mbox{Res} \, (\psi (s),s=1)
\right] \nn \\
&=& -2^{-13} \pi^{-21/2}  \left[
\sum_{k=1}^{\infty} A(k) \beta^{-2k} +G(1, \infty) \beta^{-1}
\right],
\eea
with
\beq
A(k)= (-1)^{k+1} (2^{2k}-1)\pi^{4k} \frac{ B_{2k}}{4 \, \Gamma
\left(k+\frac{1}{2} \right) \Gamma (k+1)},
\eeq
where $B_{2k}$ are the Bernoulli numbers.

Eq. (28) gives the Laurent series representation of the open
superstring free energy. This
representation is very convenient, both for low temperature and for
high-temperature calculations. In fact, the only thing to do in any
case is to take the neccessary amount of relevant terms of the
series (28). Notice, once more, that this series  is not just an
expansion: it actually provides the {\it exact} result for any
value of
$\beta$ for which the series exists.

Let us discuss this point in more detail, namely the convergence
condition
for the Laurent series (28). The usual convergence criterion (of
quocients) reads
 \beq
\lim_{k\rightarrow \infty} \frac{A(k+1)}{A(k)} \,
\beta^{-2(k+1)+2k} =
4\pi^2 \beta^{-2} \equiv q.
\eeq
For $q<1$ the series is convergent. Hence, expression (28) is
convergent
when $\beta > \beta_c =2\pi$, the Hagedorn temperature [1-4]. That
is, the Hagedorn temperature provides the convergence radius of the
above series, and vice-versa. It is interesting to notice that the
coefficients of the main part of the Laurent series (28) are
defined in the following way
\beq
A(k)= \frac{1 }{2\pi i} \oint_{\cal C} d\beta \, \beta^{k-1}
F(\beta),
\eeq
where ${\cal C}$ is an arbitrary, closed integration path interior
to the convergence circle of the series. The radius of this circle
of convergence corresponds to the critical temperature.

Now let us consider the open bosonic string. In this case  the
first expression (11) is
\bea
F_{\mbox{bosonic string}}& =&- 2^{-26} \pi^{-25/2} \frac{1}{2\pi
i} \int_{c-i\infty}^{c+i\infty} ds \,  \Gamma (s) \frac{
\Gamma \left( \frac{s}{2}-13\right)}{ \Gamma \left( \frac{s+1}{2}
\right)} \zeta (s)  \beta^{-s} \tr (M^2)^{13-s/2}, \nn \\
& & c>26,
\eea
where [17]
\beq
M^2= 2\left( \sum_{i=1}^{24} \sum_{n=1}^{\infty} nN_{ni}-1 \right).
\eeq
In the heat-kernel representation, we obtain
\bea
\tr (M^2)^{13-s/2} &=& \frac{1}{\Gamma \left( \frac{s}{2}-
13\right)} \, \int_0^{\infty} dt \, t^{s/2-14} \tr e^{-tM^2} =
 \frac{1}{\Gamma \left( \frac{s}{2}-13\right)} \, \int_0^{\infty}
dt \, t^{s/2-14} e^{2t} \nn \\
&& \times \prod_{n=1}^{\infty} \left( 1- e^{-2tn}
\right)^{-24} =
 \frac{\pi^{s/2-13}}{\Gamma \left( \frac{s}{2}-13\right)} \,
\int_0^{\infty} dt \, t^{s/2-14} \eta (it)^{-24},
\eea
where $\eta (\tau )= e^{i\pi\tau/12}  \prod_{n=1}^{\infty} \left(
1- e^{-2\pi i n\tau} \right)$ is Dedekind's eta function.

Let us consider the ultraviolet behaviour of the integrand:
\beq
 \left. \eta (it)^{-24} \right|_{t\rightarrow 0}= t^{12} e^{2\pi/t}
\left[ 1+ {\cal O} \left( e^{- 2\pi/t} \right) \right].
\eeq
The integral can be identically transformed into the following
\beq
\int_0^{\infty} dt \, t^{s/2-14} \eta (it)^{-24}=  \int_0^{\infty}
dt \, t^{s/2-14} \left[ \eta (it)^{-24}- t^{12} e^{2\pi/t} \right]
+ (-2\pi)^{s/2-1} \Gamma \left( 1- \frac{s}{2}\right).
\eeq
Thus, we get
\beq
 \tr (M^2)_{reg}^{13-s/2} = \frac{(2\pi)^{-1} \pi^{-13}}{\Gamma
\left( \frac{s}{2}-13\right)} \, \left[ 2^{s/2} \pi^s  \Gamma
\left( 1- \frac{s}{2}\right) \re (-1)^{s/2-1} + \pi^{(s-1)/2}
G(s,\mu ) \right],
\eeq
where
\beq
G(s,\mu )= 2\pi^{3/2}   \int_0^{\mu} dt \, t^{s/2-14} \left[ \eta
(it)^{-24}- t^{12} e^{2\pi/t} \right].
\eeq
In contrast with the open superstring case, the regularization in
the infrared region of the analytic part $G(s,\mu )$ (the parameter
$\mu$ provides the infrared cutoff) must be done for all Re $s$.

For the one-loop free energy we obtain
\beq
F_{\mbox{bosonic string}} =- 2^{-27} \pi^{-53/2} \frac{1}{2\pi
i} \int_{c-i\infty}^{c+i\infty} ds \, \left[ \varphi (s) + \psi (s)
\right], \ \ \ \ c>26,
\eeq
where
\bea
\varphi (s) &=& 2^{s /2 }  \mbox{Re} \,  (-1)^{\frac{s}{2} -1}
\frac{ \Gamma ( s) \Gamma \left(1- \frac{s}{2} \right)}{ \Gamma
\left( \frac{s+1}{2} \right)} \zeta (s) \left( \frac{\beta}{\pi}
\right)^{-s}, \\
\psi (s) &=& \pi^{(s-1)/2}  G(s,\mu) \,  \frac{ \Gamma ( s)}{
\Gamma \left( \frac{s+1}{2} \right)} \zeta (s)  \beta^{-s}.
\eea

The meromorphic function $\varphi (s)$ has first order poles at
$s=0$,
$s=1$ and $s=2k$, $k=1,2, \ldots$. The meromorphic function $\psi
(s)$ has first order poles at $s=1$ and $s=0$. The corresponding
residues are
\bea
\mbox{Res} \, (\varphi (s),s=2k)&=&  2^{k} \frac{ \Gamma
(2k)\zeta (2k)}{ \Gamma \left(k+ \frac{1}{2} \right) \Gamma (k)}
\left( \frac{\beta}{\pi} \right)^{-2k}, \nn \\
\mbox{Res} \, (\varphi (s),s=0)&=& \frac{1}{2\sqrt{\pi}}, \ \ \
\mbox{Res} \,
 (\psi (s),s=0) =-\frac{1}{2\pi} G(0, \mu), \nn \\
\mbox{Res} \,
(\psi (s),s=1)& =& G(1, \mu) \beta^{-1} \pi.
\eea
The pole of $\varphi$ for $s=1$  is also of first order but its
residue is imaginary.
Hence, we obtain
\beq
F_{\mbox{bosonic string}}  = -2^{-27} \pi^{-53/2} \left[
\sum_{k=1}^{\infty} A(k) x^{-2k} +G(1, \mu) \beta^{-1} +
\frac{1}{2\sqrt{\pi}}-\frac{1}{2\pi} G(0, \mu)
\right],
\eeq
where $x=\beta/\pi$ and
\beq
A(k)= 2^k \, \frac{\Gamma (2k) \zeta (2k)}{\Gamma
\left(k+\frac{1}{2} \right) \Gamma (k)}.
\eeq
Expression (43) gives the Laurent series representation for the
open bosonic string free energy. However, this free energy has an
explicit dependence on the cutoff parameter $\mu$. If we try to
remove the cutoff ($\mu \rightarrow \infty$) then, as is expected
$G(1, \mu  \rightarrow \infty) \rightarrow \infty$, and the free
energy diverges. This infrared divergence is due to the presence of
a tachyon in the bosonic string spectrum (notice that the tachyon
causes that very long cylinders are weighted with a positive
measure).

The criterion for convergence reads
 \beq
\lim_{k\rightarrow \infty} \frac{A(k+1)}{A(k)} \, \beta^{-2} \pi^2
= 8\pi^2 \beta^{-2} <1.
\eeq
Hence, $\beta > \beta_c =\pi \sqrt{8}$, the inverse of the Hagedorn
temperature.
Again, $\beta_c$ provides the convergence radius of the Laurent
series, and vice-versa.
\bs

\section{Laurent series representation for the closed bosonic
string}

In this section, we shall apply the formalism developed above to
the closed bosonic string case. As is well-known, in closed bosonic
string theory we have the mass operator [17]
\beq
M^2=4\left( \sum_{i=1}^{24}\sum_{n=1}^{\infty}  n \left( N_{n}^i +
\wt{N}_{n}^i \right)-2 \right],
\eeq
and also the constraint
\beq
 \sum_{i=1}^{24}\sum_{n=1}^{\infty}  n \left( N_{n}^i -
\wt{N}_{n}^i \right)=0.
\eeq
Incorporating the constraint (47) to (11)  with the help of the
standard identity, we get in the heat-kernel representation
\bea
&&\tr (M^2)^{13-s/2} = \frac{1}{\Gamma \left( \frac{s}{2}-
13\right)} \, \int_0^{\infty} dt \, t^{s/2-14} \tr e^{-tM^2}
\int_{-\pi}^{\pi} \frac{d\tau}{2\pi} \exp \left[ i\tau
\sum_{i=1}^{24}\sum_{n=1}^{\infty}  n \left( N_{n}^i -
\wt{N}_{n}^i \right) \right] \nn \\
&& = \frac{1}{\Gamma \left( \frac{s}{2}-13\right)} \,
\int_0^{\infty} dt \, t^{s/2-14} \int_{-\pi}^{\pi}
\frac{d\tau}{2\pi} e^{8t} \prod_{n=1}^{\infty} \left[ \left( 1-
e^{-4tn+i\tau n} \right)\left( 1- e^{-4tn-i\tau n}
\right)\right]^{-24}.
\eea
Changing variables to $\tau \rightarrow 2\tau$, $t \rightarrow
t/4$, we have
\beq
\tr (M^2)^{13-s/2} = \frac{2^{26-s}}{\Gamma \left( \frac{s}{2}-
13\right)} \, \int_0^{\infty} dt \, t^{s/2-14}  \int_{-1/2}^{1/2}
d\tau \, \left[ \eta (-\sigma ) \eta (\bar{\sigma} ) \right]^{-24},
\eeq
where $\sigma =\tau -it/(2\pi)$ and $\bar{\sigma}$ is the complex
conjugate.

The integrand is not regular at $\tau =0$, $t\rightarrow 0$. The
divergence in the infrared region is caused by the tachyon. When
$\tau =0$
\beq
 \int_0^{\infty} dt \, t^{s/2-14} \left[ \eta (-\sigma ) \eta
(\bar{\sigma} ) \right]^{-24}_{\tau =0} = (2\pi)^{s/2-13}
\int_0^{\infty} dt \, t^{s/2-14}  \eta (it)^{-48},
\eeq
and for $t\rightarrow 0$
\beq
\left. \eta (it)^{-48} \right|_{t\rightarrow 0} = t^{24} \left[ e^{4\pi
/t} + C(1)  e^{2\pi /t} + C(2) + {\cal O} \left( e^{-2\pi /t}  \right)
\right],
\eeq
where $C(1)=48$. Through simple transformations (as before, adding
and subtracting this leading behaviour), we obtain
\bea
&& \int_0^{\infty} dt \, t^{s/2-14}  \int_{-1/2}^{1/2} d\tau \,
\left[ \eta (-\sigma ) \eta (\bar{\sigma} ) \right]^{-24}
= \int_0^{\infty} dt \, t^{s/2-14} \left\{  \int_{-1/2}^{1/2} d\tau
\, \left[ \eta (-\sigma ) \eta (\bar{\sigma} ) \right]^{-24}
\right.
 \\  && \left. -
(2\pi)^{s-26} t^{24} \left[ e^{2/t} + C(1) e^{1/t} + C(2) \right]
\right\} +
  (2\pi )^{s-2} \left[ C(1) + 2^{s/2+11} \right]  \Gamma \left(
-\frac{s}{2}-11\right) (-1)^{s/2+11}. \nn
\eea
The regularized expression for the trace is
\bea
 \tr (M^2)_{reg}^{13-s/2}& =& \frac{2^{26-s}}{\Gamma \left(
\frac{s}{2}-13\right)} \, \left\{ (2\pi )^{-1} G(s,\mu ) + (2\pi
)^{s-2} \left[ C(1)+ 2^{s/2+11} \right] \right. \nn \\
 && \left. \times \Gamma \left( -
\frac{s}{2}-11\right) \re (-1)^{s/2+11} \right\},
\eea
where
\beq
G(s,\mu )= 2\pi \int_0^{\mu} dt \, t^{s/2-14} \left\{  \int_{-
1/2}^{1/2} d\tau \, \left[ \eta (-\sigma ) \eta (\bar{\sigma} )
\right]^{-24} - (2\pi)^{s-26} t^{24} \left[ e^{2/t} + C(1) e^{1/t}
+ C(2) \right] \right\}
\eeq
and $\mu$ is the cutoff parameter.

Finally, for the free energy we obtain
\bea
F_{\mbox{closed bosonic string}}& =& - 2^{-26} \pi^{-25/2}
\frac{1}{2\pi
i} \int_{c-i\infty}^{c+i\infty} ds \, \frac{ \Gamma ( s) \Gamma
\left(\frac{s}{2}-13 \right)}{ \Gamma
\left( \frac{s+1}{2} \right)} \zeta (s) \beta^{-s}  \tr
(M^2)_{reg}^{13-s/2} \nn \\
&\equiv &  - 2^{-26} \pi^{-29/2} \frac{1}{2\pi
i} \int_{c-i\infty}^{c+i\infty} ds \,
\left[ \varphi (s) + \psi (s) \right], \ \ \ \ c>26.
\eea
Here
\bea
\varphi (s) &=& \left[ C(1)+ 2^{s /2 +11} \right] \mbox{Re} \,
(-1)^{\frac{s}{2} +11}
\frac{ \Gamma ( s) \Gamma \left(- \frac{s}{2}-11 \right)}{ \Gamma
\left( \frac{s+1}{2} \right)} \zeta (s) \left( \frac{\beta}{2\pi}
\right)^{-s}, \\
\psi (s) &=& 2^{1-s} \pi  G(s,\mu) \,  \frac{ \Gamma ( s)}{
\Gamma \left( \frac{s+1}{2} \right)} \zeta (s)  \beta^{-s}.
\eea

The meromorphic function $\varphi (s)$ has first order poles at
$s=2k$, $k=1,2, \ldots$ and $s=-2k$, $k=1,2, \ldots, 11$ and a
second order pole at $s=0$. The meromorphic function $\psi
(s)$ has first order poles at $s=1$ and $s=0$. The corresponding
residues can be easily calculated:
\bea
&& \mbox{Res} \, (\varphi (s),s=0)= \frac{ C(1) +
2^{11}}{2\sqrt{\pi} 11!}  \left[ \log \left( \frac{\beta}{2\pi^2}
\right) + \frac{1}{2} \psi \left( \frac{1}{2} \right) + 2\gamma
\right] -\frac{  2^{9}}{\sqrt{\pi} 11!} \log 2,
 \nn \\
&& \mbox{Res} \, (\varphi (s),s=2k, k=1,2, \ldots )= \left(  C(1)
+ 2^{k+11} \right)  \frac{ \Gamma
(2k)\zeta (2k)}{ \Gamma \left(k+ \frac{1}{2} \right) \Gamma (k+12)}
\left( \frac{\beta}{\pi} \right)^{-2k}, \nn \\
&& \mbox{Res} \, (\varphi (s),s=-2k, k=1,2, \ldots, 11 )= \left(
C(1) + 2^{-k+11} \right)  \frac{ \Gamma
(-2k)\zeta (-2k)}{ \Gamma \left(-k+ \frac{1}{2} \right) \Gamma (12-
k)}
\left( \frac{\beta}{\pi} \right)^{2k},  \nn  \\
&& \mbox{Res} \, (\psi (s),s=1)= \left( \frac{\beta}{\pi}
\right)^{-1} G(1, \mu), \ \ \ \ \mbox{Res} \, (\psi (s),s=0)= -
\sqrt{\pi} \,  G(0, \mu),
\eea
where $\gamma$ is Euler's constant and the $\psi (1/2)$ refers to
the
digamma function ($\Gamma' /\Gamma$).

Finally, we obtain
\bea
&&F_{\mbox{closed bosonic string}} = - 2^{-2} \pi^{-29/2}
\left\{ {\sum_{k=-11}^{\infty}}' A(k) x^{-2k} + x^{-1} G(1,\mu )
\right. \nn \\
&&+ \left. \frac{ 131 \cdot 2^{3}}{\sqrt{\pi} \, 11!} \left[ \log
\left( \frac{x}{\pi} \right) + \frac{1}{2} \psi \left( \frac{1}{2}
\right) + 2\gamma \right] - \sqrt{\pi} G(0,\mu ) - \frac{195}{259}
\log 2 \right\}.
\eea
Again $x=\beta/\pi$, and the prime means that the term with $k=0$ is
missing in $\sum'$.

Expression (59) is not modular invariant, because this is not an
integral representation. The drawback of the above method for
closed strings is the use of a not so good regularization for $\tr
M^2$. For example, the fact that the regular part of the Laurent
series (59) contains a finite number of terms and the fact that the
dual symmetry is hidden in (59) (one cannot see it explicitly)
result from the regularization used above. Of course, owing to the
fact that the closed bosonic string contains infrared divergences
(i.e. tachyons) in the spectrum, the regularization which has been
used suffices for the description of $F (\beta )$ when $\beta
\rightarrow \beta^+_c$. However, in order to generalize the above
formalism to closed superstrings one ought to introduce some other
regularization.

The Hagedorn temperature can also be obtained from (59)
\beq
\lim_{k\rightarrow \infty} \frac{A(k+1)}{A(k)} \, x^{-2}= 8x^{-2}.
\eeq
Hence,  $\beta_c =\pi \sqrt{8}$. The series is convergent
absolutely inside the region $H$, where $H$ is the exterior part of
the circle $\Gamma : |x|=R$, being the radius
\beq
R=\left[ \overline{\lim}_{k \rightarrow \infty} A(k)^{1/k}
\right]^{-1}
=8.
\eeq
The series is divergent outside  $H$. Notice that the function
$F(\beta )$ is holomorphic in $H$.
\bs

\section{Discussion}

We have developed in this paper a formalism for the description of
the string free energy in terms of  a Laurent series representation. An
interesting property of this method is the fact that the Hagedorn
temperature appears as the radius of convergence of the
corresponding Laurent series. Thus, it can well lead to some new
physical interpretation of the Hagedorn temperature and of the
associated phase transition. To be remarked is also that the
Laurent series representation is exact for any value of $\beta$,
hence this representation is useful both for low- and for
high-temperature explicit estimations.

The Laurent series representation for the free energy looks quite
simple and reasonable in the case of the open (super)string,
however it is rather complicated for the bosonic string. This is an
outcome of the non-very-convenient method that we have
used to regularize the integrals which appear in the procedure. In
particular, the dual symmetry is hidden in the Laurent series
representation for the closed bosonic string. The lack of a better
suited regularization prevents us at this stage from obtaining an
immediate generalization to more complicated cases, as would be the
closed superstring and the heterotic string.

One possible way to generalize the above formalism to the closed
superstring and to the heterotic string is to employ the more convenient
regularization associated with  Wittaker functions and the
Mellin-Barns transforms of the same. This topic is currently under
investigation. Finally, we would like to point out that, even if it is,
of course, a matter of taste to use one or the other formalism in a
concrete calculation, in a given  situation where one does
 not know which of the representations is more fundamental,
one can expect that the use of different formalisms will actually add
some important information  to that obtained from just one of them.
\vspace{5mm}

\ni{\large \bf Acknowledgments}

S.D.O. wishes to thank I. Antoniadis and F. Englert for discussions
on related topics and the Particle Group at Hiroshima University
for kind hospitality.
S.D.O. has been supported by JSPS (Japan) and
E.E.  by DGICYT (Spain), research project
PB90-0022.
\bs

\appendix

\section{Appendix}

There are several ways of defining the Hagedorn temperature, one of
the simplest [3] being from the mass operator of the string
compactified on the torus. Here we will show how the asymptotic
behaviour of the level density as well as the Hagedorn temperature
can be obtained by making use of simple considerations based on the
heat kernel expansion. To start with, let us consider the field
theoretical case. The key ingredient is then the asymptotic heat
kernel expansion (see for example [18] and references therein). If
the manifold has a boundary, we have the asymptotic expansion
\beq
\tr e^{-tH} = \sum_{r=0}^{\infty} K_r t^{(r-D)/2} + A(t)
\eeq
$A(t)$ being vanishing for $t$ going to zero.
This is valid for a compact $D$-dimensional manifold, $M_c$, and
for an elliptic, second order differential operator $H$ (e.g. the
Laplace operator plus something) acting on fields defined on $M_c$.
The spectral density $\rho$ can be introduced by taking an  $M_c$  of
 large volume, so that one can write
\beq
\tr e^{-tH} = \int_0^{\infty} dE \, e^{-tE} \rho (E).
\label{ho1}
\eeq
Here $H$ plays the role of the Hamiltonian an is assumed to be
hermitian.
This asymptotic behaviour is a direct consequence of a tauberian
theorem due to Karamata, but for our purposes we give here a formal
derivation which is useful for the string case. From the identity
(for $a>0$)
\beq
t^{-a} = \frac{1}{\Gamma (a)} \int_0^{\infty} dE \, e^{-tE} E^{a-1}
\label{ho3}
\eeq
introduced into (\ref{ho1}), we get
\beq
\int_0^{\infty} dE \, e^{-tE} \left[ \rho (E) - \sum_{r=0}^{D-1}
\frac{K_r E^{(D-r)/2-1}}{\Gamma ((D-r)/2)} \right] = K_{D/2} +
{\cal O} (t) + A(t),
\eeq
and in the limit $t\rightarrow 0$ we obtain the sum rule [18]
\beq
\int_0^{\infty} dE \, \left[ \rho (E) - \sum_{r=0}^{D-1} \frac{K_r
E^{(D-r)/2-1}}{\Gamma ((D-r)/2)} \right] = K_{D/2}.
\eeq
As a consequence, the asymptotic behaviour of the spectral density
has a leading term which is given by $K_0 E^{D/2-1}$ (the Weyl
term). Notice that we do not pay attention to infrared divergences and
will be only interested in ultraviolet properties.

Let us now try to derive, along the same lines, the asymptotic
behaviour of the spectral density for the (super)string. The key
ingredients in this case are the string number operators, $N$, and
their related heat kernel expansions. It is well known that for
(super)strings in the light-cone gauge one has
\beq
\tr e^{-tN} = \sum_{n=0}^{\infty} d_n e^{-tn} = \int_0^{\infty} dx
\, e^{-tx} n(x),
\eeq
$n(x)$ being the number density.

The structure of the $N$ operator depends on the nature of the
(super)string considered. In general, it is possible to show that
its trace is exponentially divergent at the origin, so that for the
leading term one has
\beq
\tr e^{-tN} = b \, t^{-a} e^{t_c/t} + B(t)= b\sum_{k=0}^{\infty}
\frac{
t^{-a-k} t_c^k}{k!} + B(t),
\eeq
where $b$ is a constant and $B(t)$ a non-singular function. The
constants $b$ and $t_c$ and the function $B(t)$ depend on the model
under consideration. It should be noticed that there appears an
essential singularity at the origin, which may be interpreted ---from
a
geometric point of view--- as if the string hamiltonian behaved as an
effective operator acting on fields defined on an infinite
dimensional
manifold. Making use of eq. (\ref{ho3}) one arrives at
\beq
\int_0^{\infty} dx \, e^{-tx} \left[ n (x) -b \sum_{k=0}^{\infty}
\frac{t_c^k
x^{k+a-1}}{k!\Gamma (k+a)} \right] = B (t),
\eeq
which can be written as

\beq
\int_0^{\infty} dx \, e^{-tx} \left[ n (x) -b \, t_c^{(1-a)/2}
x^{(a-1)/2}I_{a-1} \left(\sqrt{t_c x} \right)\right] = B (t),
\eeq
where $I_{\nu} (z)$ is the Bessel function of imaginary argument.

{}From this relation two different results can be obtained. The first
one is a sum rule for the number density. This can be obtained in
the limit $t\rightarrow 0$, assuming that the whole singular part
in $t$ has been taken into account. The second result is the
asymptotic behaviour of the number density. Namely for large $x$,
from $I_{\nu} (z) \sim  e^z z^{-1/2}$ it follows that
\beq
n(x) \sim Cx^{a/2-3/4} e^{\sqrt{t_cx}}.
\eeq

The asymptotic behaviour of the density levels as a function of the
mass can be obtained by making use of the well known relation $x
\sim \alpha' m^2$, wherefrom it results
\beq
\rho (m) =2m\, \alpha' \ n (x(m)) \ \sim \ C m^{a-1/2}
e^{m\sqrt{t_c\alpha'}}. \label{hor}
\eeq
The physical meaning of the constant $\sqrt{t_c\alpha'} \equiv
\beta_c$ is the following: it is just the inverse of the Hagedorn
temperature. In fact, it is a limiting temperature since, if we
write down the spectral representation of the (super)string
partition function as
\beq
\tr e^{-\beta H} = \int_0^{\infty} dE \, e^{-\beta E} \rho (E)
\eeq
and make use of the asymptotic behaviour given by eq. (\ref{hor}),
we see that the above integral converges as soon as $\beta
>\beta_c$. Furthermore, near the critical value, we have
\beq
\tr e^{-\beta H} \sim \frac{C\Gamma (a+1/2)}{(\beta-
\beta_c)^{a+1/2}}.
\eeq
In consequence, the free energy has a logarithmic singularity,
namely
\beq
F_{\beta} =- \frac{1}{\beta} \log \tr e^{-\beta H} =
\frac{1}{\beta} \left( a+\frac{1}{2} \right) \log (\beta-\beta_c)
+ \mbox{const.}
\eeq
It must be pointed out, however, that the physical meaning of such
limiting temperature and that of the associated phase transition are
still lacking [3].
\bs

\section{Appendix}

In this appendix we are going to discuss the issue of the
analytical continuation of the free energy for open (super)strings
above the Hagedorn temperature.

The starting point will be our explicit Laurent  series
representation for the open (super)string free energy.  In
particular, let us consider the series (see (28) and (43)):
\beq
f(x)=\sum_{k=1}^{\infty} A(k) x^{-k}, \label{b1}
\eeq
where  $x=(\beta /\pi )^2$ and
\beq
A(k)= \frac{ \Gamma
(2k)\zeta (2k)}{ \Gamma \left(k+ \frac{1}{2} \right) \Gamma (k)}
\times \left\{ \brr{ll} 1-2^{-k}, & \mbox{for the open
superstring,} \\ 2^{k}, & \mbox{for the open bosonic string.} \err
\right.
\eeq
The radius of convergence of these series is
\beq
R=\left[ \overline{\lim}_{k \rightarrow \infty} A(k)^{1/k}
\right]^{-1}=\frac{1}{\rho} = \left\{ \brr{ll} \frac{1}{4}, &
\mbox{for superstrings,} \\  \frac{1}{8}, & \mbox{for  bosonic
strings.} \err \right.
\eeq
{}From the discussion carried out in the paper we see that the
condition $Rx>1$ translates, for bosonic strings, into $\beta
>\beta_c =\pi \sqrt{8}$, and for superstrings into $\beta >\beta_c
=2\pi$.

We can now proceed with the construction of the analytic
continuation of $f(x)$ to values $x\leq \rho$, i.e., to $\beta \leq
\beta_c$ (that is, above the Hagedorn temperature). In order to
sum the divergent series for  $x\leq \rho$ we will use the Borel
procedure. We define the function $g(x)$, connected to $f(x)$ by a
Borel transform for any $\varepsilon >0$:
\bea
g(x) &=& \frac{1}{2\pi i} \oint_{|t|=\rho + \varepsilon} dt \, f(t)
e^{tx} = \frac{1}{2\pi i} \oint_{|t|=\rho + \varepsilon} dt \,
\sum_{k=1}^{\infty} A(k) t^{-k} e^{tx} \nn \\
&=& \sum_{k=1}^{\infty}  \frac{A(k)}{2\pi i} \oint_{|t|=\rho +
\varepsilon} dt \,  t^{-k} e^{tx} =  \sum_{k=1}^{\infty}
\frac{A(k)}{(k-1)!} x^{k-1}.
\eea
Hence,
\beq
f(x)= \int_{0}^{\infty} dt \, g(t) e^{-tx} =  \int_{0}^{\infty} dt
\, \sum_{k=1}^{\infty}  \frac{A(k)}{(k-1)!} t^{k-1} e^{-tx}.
\label{abh}
\eeq
In this way, using the Borel summation formula we have been able
to construct the analytic continuation of the open (super)string
free energy (expression (\ref{abh})) to temperatures above the
Hagedorn one ( $\beta \leq \beta_c$)\, !. Indeed, the resulting
expression for $f(x)$ is convergent for any $x$, i.e. for any
$\beta$. This is a remarkable advantage of our representation for
the free energy in terms of a Laurent series as compared with the
standard, integral representation associated to the Jacobi theta
functions: it provides a natural way to analitically continuing the
free energy above the Hagedorn temperature and hence, perhaps, to
the explicit construction of the free energy operator corresponding
to the new string phase. Of course, the precise interpretation of
the free energy obtained in this way is somehow delicate and many
questions can be asked here. In particular, does it actually
describe a new phase of string theory at non-zero temperature? or,
how is it related with the known models (topological strings) which
also pretend to play the role of the new string phase?. These
questions deserve further study.
\bs

\section{Appendix}

In this appendix a way to obtain a simple estimation of the
coefficients of the Laurent series for the free energy
corresponding  the open (super)string case is described.
Let us consider again the series (\ref{b1}).

All the coefficents $A(k)$ are real and positive, hence we can use
Pringsheim's theorem: the point $x=\rho$, which is located on the
border of the convergence circle corresponding to the power series
$f(x)$, is not a regular point. Moreover, since $f(x=\rho) =
\infty$, this point is a pole of $f(x)$.

Let $Z(k)\equiv A(k)^{1/k}$ and $\overline{\lim}_{k\rightarrow
\infty} Z(k) =\rho$. Then starting from a value of $N$ big enough,
we can write $Z(N)\simeq \rho$, and the series is approximated by
\beq
f(x)= \sum_{k=1}^{\infty} \left( \frac{Z(k)}{x} \right)^k \simeq
\sum_{k=1}^{N-1} A(k) x^{-k} +  \sum_{k=N}^{\infty} \left(
\frac{\rho}{x} \right)^k =\sum_{k=1}^{N-1} A(k) x^{-k} +   \left(
\frac{\rho}{x} \right)^N \frac{x}{x-\rho}.
\eeq
For fixed $N$, we can expand $x^{-N+1}$ in powers of $x-\rho$
\beq
x^{-N+1}=\rho^{-N+1}- (N-1)\rho^{-N} (x-\rho) + {\cal O} \left( (x-
\rho )^2 \right),
\eeq
and then
\beq
\rho^N x^{-N+1} \frac{1}{x-\rho} = \frac{\rho}{x-\rho} -(N-1)+
{\cal O} ( (x-\rho ).
\eeq
Finally,
\beq
f(x)= \sum_{k=1}^{\infty}  A(k) x^{-k} \simeq \sum_{k=1}^{N-1} A(k)
x^{-k} +  \frac{\rho}{x-\rho} -(N-1).
\eeq
This expression gives an asymptotic estimate for the free energy
at high temperature (near the Hagedorn temperature). The pole at
$\beta = \beta_c$ is explicitly written as the second term on the r.h.s.
We thus see that the Laurent series representation of the free energy
is very useful for the high temperature expansion. In the same way
one can discuss the low energy expansion.
\newpage



\end{document}